\documentclass[12pt,letterpaper]{article}

\usepackage[margin=1in]{geometry}
\usepackage{amsmath,amssymb,bm}
\usepackage{graphicx}
\usepackage{natbib}
\usepackage{booktabs}
\usepackage{enumitem}
\usepackage{setspace}
\usepackage[hidelinks]{hyperref}
\usepackage{xcolor}
\usepackage[export]{adjustbox}

\newcommand{\bbR}{\mathbb{R}}
\newcommand{\bA}{\mathbf{A}}
\newcommand{\bC}{\mathbf{C}}

\newcommand{\bS}{\mathbf{S}}
\newcommand{\bV}{\mathbf{V}}
\newcommand{\bW}{\mathbf{W}}

\newcommand{\br}{\mathbf{r}}
\newcommand{\bbeta}{\bm{\beta}}
\newcommand{\bdelta}{\bm{\delta}}
\newcommand{\beps}{\bm{\varepsilon}}
\newcommand{\btheta}{\bm{\theta}}
\newcommand{\bSigma}{\bm{\Sigma}}

\newcommand{\bTheta}{\bm{\Theta}}

\title{Fusing Sparse Observations and Dense Simulations for Spatial Extreme Value Analysis: Application to U.S.\ Coastal Sea Levels}

\author{
Brian N. White\textsuperscript{1,*}, Brian Blanton\textsuperscript{2}, Rick Luettich\textsuperscript{2,3}, Richard L. Smith\textsuperscript{1} \\[6pt]
{\normalsize \textsuperscript{1}Department of Statistics and Operations Research, University of North Carolina at Chapel Hill} \\
{\normalsize \textsuperscript{2}Renaissance Computing Institute (RENCI), University of North Carolina at Chapel Hill} \\
{\normalsize \textsuperscript{3}Institute of Marine Sciences, University of North Carolina at Chapel Hill} \\[6pt]
{\normalsize \textsuperscript{*}Correspondence: bnw@unc.edu}
}

\date{}

\begin{document}

\maketitle

\begin{abstract}
Estimating spatial extremes from sparse observational networks produces
uncertain return level maps, but dense output from physics-based simulation
models is often available as a complementary data source. We develop a
two-stage frequentist framework for fusing observations and simulations. In
Stage~1, generalized extreme value (GEV) distributions are fitted
independently at each site, with a nonstationary location parameter where
appropriate to accommodate observed trends. In Stage~2, the parameter
estimates from all sources are modeled jointly as a high-dimensional spatial
process through a linear model of coregionalization (LMC). Cross-source
correlations, estimated from spatially interspersed networks without
co-located sites, provide the mechanism for information transfer; an
analytic gradient for the resulting likelihood keeps estimation
computationally practical. We apply the framework to U.S.\ coastal sea
levels over 1979--2021, fusing 29 NOAA tide gauge records with 100 ADCIRC
hydrodynamic simulation sites. Leave-one-out cross-validation shows a 35\%
reduction in 100-year return level RMSE relative to a gauge-only model.
Geographic block cross-validation confirms that fusion benefits persist
under spatial extrapolation. The approach is implemented in the R package
\texttt{evfuse}.
\end{abstract}


\section{Introduction}
\label{sec:intro}

Extreme value distributions vary in space, but observational networks are
seldom dense enough to estimate this variation precisely. Physics-based
simulation models can produce the quantity of interest at far greater
spatial density, and fusing their output with sparse observations can
improve both the coverage and precision of return level estimates, the
$T$-year thresholds expected to be exceeded on average once every $T$
years. We develop a general framework for
this fusion and demonstrate it by combining tidal gauge observations with
hydrodynamic simulation output to map extreme coastal sea levels along the
U.S.\ East and Gulf Coasts.

The two data sources exemplify a structure common in environmental
monitoring. The NOAA tidal gauge network provides reliable long-term sea
level records, but gauges are sparse and their records are short relative to
the return periods of interest, producing substantial uncertainty in tail
parameter estimates at individual sites. Physics-based hydrodynamic models
such as ADCIRC \citep{Luettich1992} can simulate sea levels at hundreds of
coastal locations, providing the spatial density that gauges lack, but their
output reflects model physics rather than direct observation.

Statistical approaches to spatial extremes include max-stable processes and hierarchical GEV models with spatially varying parameters \citep{Davison2012, Cooley2007, Sang2010}. Max-stable models are theoretically appealing and capture the spatial extent of extreme events, but their likelihood is intractable for even moderate numbers of stations. When the primary interest is marginal return levels rather than joint tail dependence, latent process models are often preferred: the GEV parameters at each location are treated as realizations of a spatial process, and inference proceeds either through Bayesian hierarchical modeling \citep{Cooley2007, Sang2010, Gelfand2004, Finley2008} or through a two-stage frequentist approach where pointwise MLEs are smoothed spatially \citep{Russell2020}. Data fusion combining in-situ observations with gridded model output or remote sensing has a rich history in spatial statistics \citep{Cressie2011, Nguyen2012, Berliner2003}. Despite this, fusion methods have rarely been applied to extreme value parameters, where the goal is to combine heterogeneous sources to improve tail inference rather than to interpolate mean fields.

We adopt a two-stage frequentist approach for computational tractability and interpretability, building on \citet{Russell2020}, who combined a linear model of coregionalization (LMC) with two-stage frequentist inference for spatially varying GEV parameters, applied to a single network of Gulf Coast precipitation stations. Our contribution extends this framework in two directions.

First, we generalize the model to represent GEV parameters from multiple data sources as a single high-dimensional spatial process. The LMC is identifiable under heterotopic sampling \citep{Wackernagel2003}, so cross-source correlations can be estimated from spatially interspersed networks without co-located sites. A selection matrix extracts the observed components from the full latent covariance, yielding a standard Gaussian likelihood. An analytic gradient for the resulting 33-parameter log-likelihood, derived from the LMC structure (Section~S1), reduces optimization from approximately 90 minutes to under 3 minutes on a desktop workstation, making multi-start optimization and bootstrap inference practical.

Second, we provide cross-validation evidence quantifying the gain from fusion. Under a nonstationary Stage~1 model that accommodates observed trends, the joint model reduces 100-year return level LOO-CV RMSE by 35\% relative to a NOAA-only baseline. Geographic block cross-validation shows that the benefit persists under spatial extrapolation, though at reduced magnitude.

\section{Data}
\label{sec:data}

The spatial domain extends from southern Texas ($26.1^\circ$\,N, $97.2^\circ$\,W) to northern Maine ($44.9^\circ$\,N, $67.0^\circ$\,W). We analyze annual maximum sea levels over 1979--2021 ($T_0 = 43$ years) from two sources.

\subsection{NOAA Tidal Gauge Observations}

We obtained hourly sea-level measurements from the NOAA Tides \& Currents portal in May 2022. All measurements are in meters relative to the mean sea level (MSL) tidal datum (1983--2001 epoch).

We applied a two-step quality filter: (i) any year with fewer than 90\% of expected hourly observations was excluded; (ii) any station with fewer than 20 annual maxima after step~(i) was omitted. This yielded $L_N = 29$ stations, each with at least 34 annual maxima (24 of 29 exceeding~40). The 90\% threshold guards against missing the true annual maximum. These records are short relative to the return periods of interest (100 years): estimating the shape parameter~$\xi$, which controls tail behavior, from 34--43 annual maxima produces substantial uncertainty at individual sites, motivating spatial borrowing of strength.

If tidal gauges fail during extreme events, the resulting censorship of the largest observations would bias $\hat{\xi}$ downward. This provides additional motivation for fusion: simulations are not subject to instrument failure, and the cross-source shape correlation in Stage~2 can mitigate such bias.

\subsection{ADCIRC Simulated Data}

The Advanced Circulation (ADCIRC) model is a finite-element hydrodynamic model that solves the shallow water equations on unstructured grids \citep{Luettich1992, Dietrich2011}. Simulated hourly sea-level values over 1979--2021 at $L_A = 100$ coastal sites were generated by the Renaissance Computing Institute (RENCI) at the University of North Carolina at Chapel Hill, driven by CFSR reanalysis atmospheric forcing. These are historical reconstructions generated from model physics alone. Although ADCIRC operates on unstructured grids with thousands of nodes, we use 100 representative coastal sites, a density sufficient to resolve the large-scale spatial patterns of interest along 6{,}000\,km of coastline. Finer-resolution questions would require a denser draw from the ADCIRC grid and corresponding extensions to the spatial model (Section~\ref{sec:limitations}). Because the record is complete by design, all 43 annual maxima are available at each site.

\subsection{Study Design}

\begin{figure}[!htbp]
\centering
\includegraphics[width=\textwidth]{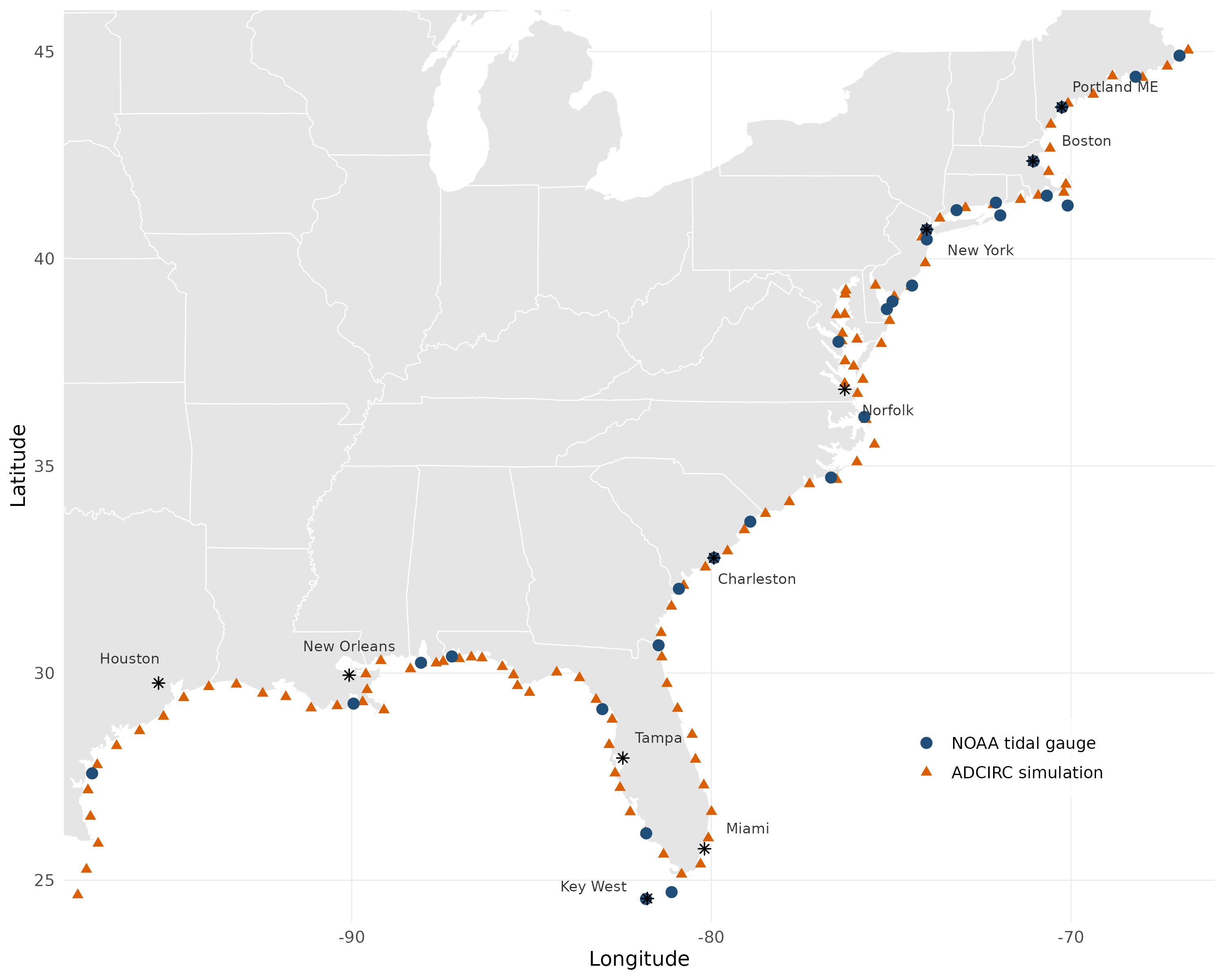}
\caption{Study area showing the locations of 29 NOAA tidal gauge stations (circles) and 100 ADCIRC simulation sites (triangles) along the U.S.\ East and Gulf Coasts.}
\label{fig:study_area}
\end{figure}

The total number of sites is $L = L_N + L_A = 129$, with no site having both NOAA and ADCIRC observations. The ADCIRC sites fill spatial gaps in the NOAA network, particularly along the Gulf Coast and southeastern Atlantic (Figure~\ref{fig:study_area}). The median nearest-neighbor distance between a NOAA and an ADCIRC site is 21.7\,km (range: 1.9--119.8\,km), providing dense intermixing of the two networks.

\subsection{Trend Diagnostics}

Before fitting the spatial model, we assess the temporal stationarity of annual maxima at each site using the Mann-Kendall nonparametric trend test \citep{Mann1945, Kendall1975} with Sen's slope estimator \citep{Sen1968}. At NOAA sites, 19 of 29 (66\%) show significant positive trends ($p < 0.05$), with a median Sen's slope of 4.6\,mm/yr. By contrast, only 1 of 100 ADCIRC sites shows a significant trend (as expected at $\alpha = 0.05$), and the median ADCIRC slope (0.4\,mm/yr) is an order of magnitude smaller (Table~\ref{tab:trends}; Figure~S2). ADCIRC's near-zero trends are expected: the simulations are driven by CFSR reanalysis forcing that does not incorporate long-term sea level rise, so any trends reflect only interannual variability in atmospheric forcing.

The trends are modest relative to return levels: the median slope of 4.6\,mm/yr produces a shift of roughly 0.2\,m over the 43-year record, compared to 100-year return levels of 1--4\,m. Nevertheless, this asymmetry between trending NOAA and stationary ADCIRC affects the cross-source correlations estimated in Stage~2, because NOAA's annual maxima contain trend-driven variance that ADCIRC's simulations lack.

\section{Statistical Framework}
\label{sec:framework}

\subsection{Stage 1: Marginal GEV Fitting}
\label{sec:stage1}

Several well-established methods exist for modeling extreme values from environmental time series \citep{Coles2001}, including annual block maxima via the generalized extreme value (GEV) distribution and peaks-over-threshold via the generalized Pareto distribution. We adopt block maxima here. At each site $s_l$ ($l = 1, \ldots, L$), we fit a GEV distribution to the annual maxima by maximum likelihood \citep[see][Ch.~4]{Coles2001}, avoiding threshold selection across 129 heterogeneous sites and ensuring compatibility with the two-source design. The GEV distribution function is
\begin{equation}
F(y \mid \mu, \sigma, \xi) = \exp\left\{-\left[1 + \xi\left(\frac{y - \mu}{\sigma}\right)\right]^{-1/\xi}\right\},
\label{eq:gev}
\end{equation}
defined on $\{y : 1 + \xi(y - \mu)/\sigma > 0\}$, with location $\mu \in \bbR$, scale $\sigma > 0$, and shape $\xi \in \bbR$ \citep{Coles2001}. The $T$-year return level is
\begin{equation}
r_T = \mu + \frac{\sigma}{\xi}\bigl[(-\log(1 - 1/T))^{-\xi} - 1\bigr].
\label{eq:rl}
\end{equation}

Because 66\% of NOAA sites exhibit significant positive trends (Section~\ref{sec:data}), we fit a nonstationary GEV at NOAA sites with $\mu(t) = \mu_0 + \mu_1(t - t_{\text{ref}})$, $t_{\text{ref}} = 2000$, and $\sigma, \xi$ stationary. Centering at $t_{\text{ref}} = 2000$ (near the record midpoint) makes $\hat{\mu}_0$ and $\hat{\mu}_1$ approximately orthogonal, minimizing the intercept's standard error. Under this centering, $\mu_0$ is the location parameter at year 2000, and return levels have a concrete temporal interpretation: the $T$-year return level represents the water level exceeded with probability $1/T$ per year under year-2000 conditions. ADCIRC sites, which show no significant trends, retain stationary fits.

The output is a triplet $\hat{\btheta}(s_l) \in \bbR^3$ at each site: $(\hat{\mu}_0, \log\hat{\sigma}, \hat{\xi})^\top$ at NOAA sites (year-2000 intercept) and $(\hat{\mu}, \log\hat{\sigma}, \hat{\xi})^\top$ at ADCIRC sites (stationary MLE). The trend slope $\mu_1$ is excluded from the spatial model because ADCIRC sites have no trend to co-model. Including $\mu_1$ as a seventh dimension would require either fixing its cross-source correlation to zero or estimating it from NOAA sites alone; neither adds spatial information beyond what $\mu_0$ already provides. A seven-dimensional extension incorporating $\mu_1$ is a natural direction for climate projection applications. The log-scale parameterization ensures unconstrained optimization in Stage~2; in other applications, different link functions could be used for other parameter types, but the statistical framework is unchanged.

\subsection{Stage 2: Multivariate Spatial Model}
\label{sec:stage2}

\subsubsection{Latent Process and Coregionalization}
\label{sec:lmc}

The LMC assumes that the six GEV parameters at any location are linear combinations of six independent spatial patterns, each with its own correlation range, so that different parameter combinations can exhibit different scales of spatial variation. At each location $s \in D$, define the six-dimensional latent vector
\begin{equation}
\btheta(s) = \bigl(\underbrace{\mu_N(s),\; \log\sigma_N(s),\; \xi_N(s)}_{\text{NOAA}},\; \underbrace{\mu_A(s),\; \log\sigma_A(s),\; \xi_A(s)}_{\text{ADCIRC}}\bigr)^\top.
\label{eq:theta6}
\end{equation}
Under the nonstationary Stage~1 model, $\mu_N(s)$ denotes the year-2000 intercept $\mu_0(s)$ at NOAA sites, while $\mu_A(s)$ is the stationary location at ADCIRC sites. Because ADCIRC shows no trends and the NOAA reference year ($t_{\text{ref}} = 2000$) lies near the record midpoint, the two quantities are approximately comparable as location parameters at a common epoch. Both NOAA and ADCIRC GEV parameters exist as latent quantities at every location, even though only one source is observed at any given site. We model $\btheta(s)$ via the linear model of coregionalization (LMC; \citealt{Journel1978}; \citealt{Wackernagel2003}; see also \citealt{Finley2008}):
\begin{equation}
\btheta(s) = \bbeta + \bA\,\bdelta(s),
\label{eq:lmc}
\end{equation}
where $\bbeta \in \bbR^6$ is the spatial mean, $\bA$ is a $6 \times 6$ lower triangular matrix, and $\bdelta(s) = (\delta_1(s), \ldots, \delta_6(s))^\top$ consists of six independent, mean-zero Gaussian processes with exponential covariance $\mathrm{Cov}(\delta_i(s), \delta_i(s')) = \exp(-\|s - s'\|/\rho_i)$, $\rho_i > 0$. The exponential covariance corresponds to a Mat\'ern model with smoothness $\nu = 1/2$ \citep[see][]{Stein1999}, chosen for parsimony given that 129 sites provide limited power to distinguish smoothness. Throughout, $\|s - s'\|$ denotes great circle (haversine) distance in kilometers.

The marginal covariance at any location is $\bA\bA^\top$, whose off-diagonal blocks encode cross-source correlations, which are the quantities enabling information transfer. The cross-covariance between two locations $s$ and $s'$ is
\begin{equation}
\mathrm{Cov}(\btheta(s),\; \btheta(s')) = \bA\,\mathrm{diag}\bigl(\exp(-\|s-s'\|/\rho_1),\;\ldots,\;\exp(-\|s-s'\|/\rho_6)\bigr)\,\bA^\top.
\label{eq:cov}
\end{equation}

Because each latent GP $\delta_i$ has its own range parameter $\rho_i$, different linear combinations of the six GEV parameters can exhibit different spatial correlation lengths. For example, if the first GP has a large range $\rho_1$ and loads primarily onto $\mu_N$ and $\mu_A$ (via $A_{1,1}$ and $A_{4,1}$), the location parameters will be spatially smooth while other parameter combinations vary over shorter distances. A separable model with a single shared range parameter cannot capture this structure.

\textit{Identifiability.}\quad A natural question is whether cross-source covariance parameters are identifiable without co-located observations. Under the LMC, they are: the shared range parameters $\rho_k$ are identified from within-source pairs, and the cross-source coefficients are then determined by a linear system evaluated at cross-source distances \citep[Ch.~23]{Wackernagel2003}. This requires distinct range parameters and sufficient spatial intermixing of the two networks, both satisfied here, as the median inter-source distance (21.7\,km) is far shorter than the shortest estimated range (188\,km).

\subsubsection{Partial Observations and the Selection Matrix}
\label{sec:selection}

At observed sites, $\hat{\btheta}(s_l) = \btheta(s_l) + \beps(s_l)$, where $\beps(s_l)$ is Stage~1 estimation error. At NOAA sites, only components 1--3 of $\btheta(s_l)$ are observed; at ADCIRC sites, only components 4--6. Let $n_{\text{obs}} = 3L_N + 3L_A = 387$. Each observation $m$ is associated with a parameter index $p(m) \in \{1,\ldots,6\}$ and a site index $s(m) \in \{1,\ldots,L\}$. The $n_{\text{obs}} \times 6L$ selection matrix $\bS$ extracts observed entries from the full latent vector $\btheta_{\text{full}} \in \bbR^{6L}$ (stored in parameter-major order). It has exactly one nonzero entry per row:
\begin{equation}
S_{m,\,(p(m)-1)L + s(m)} = 1, \qquad \text{all other entries zero.}
\label{eq:selection}
\end{equation}
In plain terms, each row of $\bS$ picks out one observed parameter at one site from the full latent vector. This formalism accommodates any pattern of partial observations: sites with one source, multiple sources, or different subsets of parameters.

\subsubsection{Likelihood}
\label{sec:likelihood}

The observed parameter estimates $\hat{\bTheta}_{\text{obs}}$ are noisy versions of the true latent parameters. Their covariance $\bV$ therefore has two components: a signal term $\bSigma_{\text{obs}}$ from the coregionalization model, representing genuine spatial structure, and a noise term $\bW$ from Stage~1 estimation error. The matrix $\bW$ is estimated once from the bootstrap (described below) and held fixed throughout Stage~2 optimization, so that only the coregionalization parameters $(\bbeta, \bA, \bm{\rho})$ are estimated by maximum likelihood.

Under the model:
\begin{equation}
\hat{\bTheta}_{\text{obs}} \sim \mathcal{N}\bigl(\bS(\bbeta \otimes \mathbf{1}_L),\;\bV\bigr), \qquad \bV = \bSigma_{\text{obs}} + \bW,
\label{eq:gaussian}
\end{equation}
where $\hat{\bTheta}_{\text{obs}} \in \bbR^{n_{\text{obs}}}$ is the vector of Stage~1 MLEs, stacked in the same parameter-major order as $\btheta_{\text{full}}$: all $L$ location estimates first, then all $L$ log-scale estimates, then all $L$ shape estimates, with each site contributing the three parameters from whichever source is observed. The mean vector $\bbeta = (\beta_1, \ldots, \beta_6)^\top$ follows the ordering of~\eqref{eq:theta6}: $\beta_1 = \mathbb{E}[\mu_N]$, $\beta_2 = \mathbb{E}[\log\sigma_N]$, $\beta_3 = \mathbb{E}[\xi_N]$, $\beta_4 = \mathbb{E}[\mu_A]$, $\beta_5 = \mathbb{E}[\log\sigma_A]$, $\beta_6 = \mathbb{E}[\xi_A]$, and the lower triangular entries of $\bA$ and the range parameters $\rho_1, \ldots, \rho_6$ inherit the same ordering. The observed coregionalization covariance is $\bSigma_{\text{obs}} = \bS\bSigma\bS^\top$. For observations $m, n$ with parameter indices $p(m), p(n)$ and site indices $s(m), s(n)$:
\begin{equation}
[\bSigma_{\text{obs}}]_{mn} = \sum_{i=1}^{6} A_{p(m),i}\,A_{p(n),i}\,\exp(-\|s_{s(m)} - s_{s(n)}\|/\rho_i),
\label{eq:sigma_obs}
\end{equation}
computed in $O(n_{\text{obs}}^2 \cdot 6)$ without forming the full $6L \times 6L$ matrix.

Letting $\br = \hat{\bTheta}_{\text{obs}} - \bS(\bbeta \otimes \mathbf{1}_L)$, the negative log-likelihood to be minimized over $(\bbeta, \bA, \bm{\rho})$ is
\begin{equation}
\ell(\bbeta, \bA, \bm{\rho}) = \tfrac{1}{2}\bigl[n_{\text{obs}}\log(2\pi) + \log|\bV| + \br^\top\bV^{-1}\br\bigr],
\label{eq:nll}
\end{equation}
where $\bV$ depends on $(\bA, \bm{\rho})$ through~\eqref{eq:sigma_obs} and the fixed $\bW$. The optimization is over the 33-dimensional parameter vector $(\bbeta, \mathrm{vech}(\bA), \log\bm{\rho}) \in \bbR^{33}$, where $\mathrm{vech}(\bA)$ denotes the 21 lower-triangular entries of $\bA$ and the log-rho reparameterization enforces positivity of the range parameters.

The measurement error covariance $\bW$ is estimated via a nonparametric block bootstrap ($B = 500$ replicates), resampling years jointly across all sites within each source to preserve cross-site storm dependence. Years are resampled independently across sources because the Stage~1 estimation errors arise from distinct data-generating mechanisms. In finite samples, shared temporal structure could induce residual cross-source dependence; a joint resampling scheme would provide a conservative alternative.

The bootstrap covariance is regularized by Wendland covariance tapering \citep{Wendland1995, Furrer2006}, following \citet{Russell2020}, with range $\lambda = 300$\,km. The Wendland $C_4$ function $\psi(d/\lambda) = (1 - d/\lambda)_+^6(35(d/\lambda)^2 + 18\,d/\lambda + 3)/3$ is a compactly supported positive-definite radial basis function that equals zero for $d \geq \lambda$. Each entry of $\bW$ is multiplied by $\psi(d_{ij}/\lambda)$, suppressing spurious long-range correlations from the finite bootstrap. To select $\lambda$, we binned site pairs by distance and computed mean absolute bootstrap correlations per bin. Correlations decay from $\sim$0.6 at 0--50\,km to $\sim$0.2 at 200--500\,km, then flatten to a noise floor of $\sim$0.10. The value $\lambda = 300$\,km captures the real signal while tapering to zero before the noise floor. Results are negligibly sensitive to this choice over $\lambda \in \{150, 300, 500\}$\,km (Table~\ref{tab:taper_sensitivity}).

\subsection{Optimization}
\label{sec:gradient}

The negative log-likelihood~\eqref{eq:nll} is minimized using L-BFGS-B with analytic gradients derived from the LMC structure (Section~S1). Twenty random restarts from log-uniform$[50, 5000]$\,km initial ranges guard against local optima.

\subsection{Kriging with Partial Observations}
\label{sec:kriging}

Given fitted parameters $(\tilde{\bbeta}, \tilde{\bA}, \tilde{\bm{\rho}})$, the kriging mean and covariance at a new location $s_0$ are
\begin{equation}
\tilde{\btheta}(s_0) = \bbeta + \bC^\top\bV^{-1}\br,
\label{eq:krig_mean}
\end{equation}
\begin{equation}
\bV_{\btheta}(s_0) = \bA\bA^\top - \bC^\top\bV^{-1}\bC,
\label{eq:krig_cov}
\end{equation}
where $C_{m,k} = \sum_{i=1}^{6} A_{p(m),i}\,A_{k,i}\,\exp(-\|s_{s(m)} - s_0\|/\rho_i)$. The first three components of $\tilde{\btheta}(s_0)$ give the predicted NOAA parameters $(\tilde{\mu}_N, \log\tilde{\sigma}_N, \tilde{\xi}_N)$, and the upper-left $3 \times 3$ block of $\bV_{\btheta}(s_0)$ gives their covariance. The $T$-year return level at $s_0$ is $\tilde{r}_T(s_0) = g(\tilde{\mu}_N, \log\tilde{\sigma}_N, \tilde{\xi}_N)$ via~\eqref{eq:rl}, with standard error obtained by the delta method:
\begin{equation}
\mathrm{SE}(\tilde{r}_T) = \sqrt{\nabla g^\top\,[\bV_{\btheta}(s_0)]_{1:3,\,1:3}\,\nabla g},
\label{eq:delta_se}
\end{equation}
where $\nabla g = (\partial r_T/\partial\mu,\;\partial r_T/\partial\log\sigma,\;\partial r_T/\partial\xi)^\top$. Letting $y_T = -\log(1 - 1/T)$, the partial derivatives are
\begin{equation}
\frac{\partial r_T}{\partial\mu} = 1, \qquad \frac{\partial r_T}{\partial\log\sigma} = r_T - \mu, \qquad \frac{\partial r_T}{\partial\xi} = \frac{\sigma}{\xi^2}\bigl(1 - y_T^{\xi}\bigr) - \frac{\sigma}{\xi}\,y_T^{-\xi}\log y_T.
\label{eq:rl_grad}
\end{equation}
For the results reported below, we compute standard errors by Monte Carlo simulation (10{,}000 draws from the kriging posterior), which correctly propagates the nonlinearity of~\eqref{eq:rl}. The delta method provides a useful closed-form check; a comparison of the two approaches is given in Section~\ref{sec:uncertainty_components} and Figure~S6.

\subsection{Model Assessment}
\label{sec:assessment}

We compare the joint model against NOAA-only and ADCIRC-only baselines (each a 3-dimensional LMC). The joint model has 33 free parameters over 129 sites (6 mean, 21 coregionalization, 6 range); each single-source baseline has 12 parameters. We evaluate predictive performance through two cross-validation designs of increasing difficulty, followed by sensitivity analyses.

LOO-CV at the 29 NOAA sites uses a closed-form block shortcut that avoids refitting the model 29 times, conditioning on the full-data covariance parameter estimates \citep{Rasmussen2006}. Because each NOAA site contributes a block of three observations ($\mu, \log\sigma, \xi$), we use the block generalization (Section~S1). We report LOO-RMSE for each parameter, for 100-year return levels, total log predictive density (LPD), and calibration via probability integral transform (PIT) histograms and empirical coverage of predictive intervals.

Geographic block cross-validation tests fusion under spatial extrapolation by holding out five contiguous coastal blocks in turn (Gulf Coast, $n = 4$; Florida, $n = 5$; Southeast, $n = 5$; Mid-Atlantic, $n = 6$; New England, $n = 9$), retaining all ADCIRC sites and refitting Stage~2 from scratch for both models in each fold.

We assess robustness to three analyst choices: (i) simulation network density, by subsampling ADCIRC sites; (ii) bootstrap taper range $\lambda$; and (iii) bootstrap batch variation. A simulation study further confirms that the cross-source correlations are recoverable under the observed non-co-located sampling design (Section~\ref{sec:identifiability}).

\section{Results}
\label{sec:results}

\subsection{Marginal GEV Fits}
\label{sec:stage1_results}

All 129 GEV fits converge. Anderson-Darling goodness-of-fit tests reject the GEV null at $\alpha = 0.05$ for only 1 of 129 sites ($p = 0.043$), consistent with the $\sim$6.5 false positives expected by chance; no site is significant after Bonferroni correction. QQ plots and bootstrap sampling distributions confirm adequate fit at all sites (Section~S2).
Bootstrap standard errors for $\hat{\xi}$ are substantial (median 0.21 at NOAA, 0.17 at ADCIRC), a direct consequence of the short records and the primary motivation for spatial smoothing in Stage~2. The two-stage structure makes the Gaussian approximation in Stage~2 directly verifiable: bootstrap sampling distributions for $\hat{\xi}$ show mild positive skew at the heaviest-tailed Gulf Coast stations but no substantial departures from normality (Figure~S3), and PIT calibration (Table~\ref{tab:calibration}) confirms that this approximation does not produce miscalibrated predictions.

\begin{figure}[!htbp]
\centering
\includegraphics[width=7.5in,center]{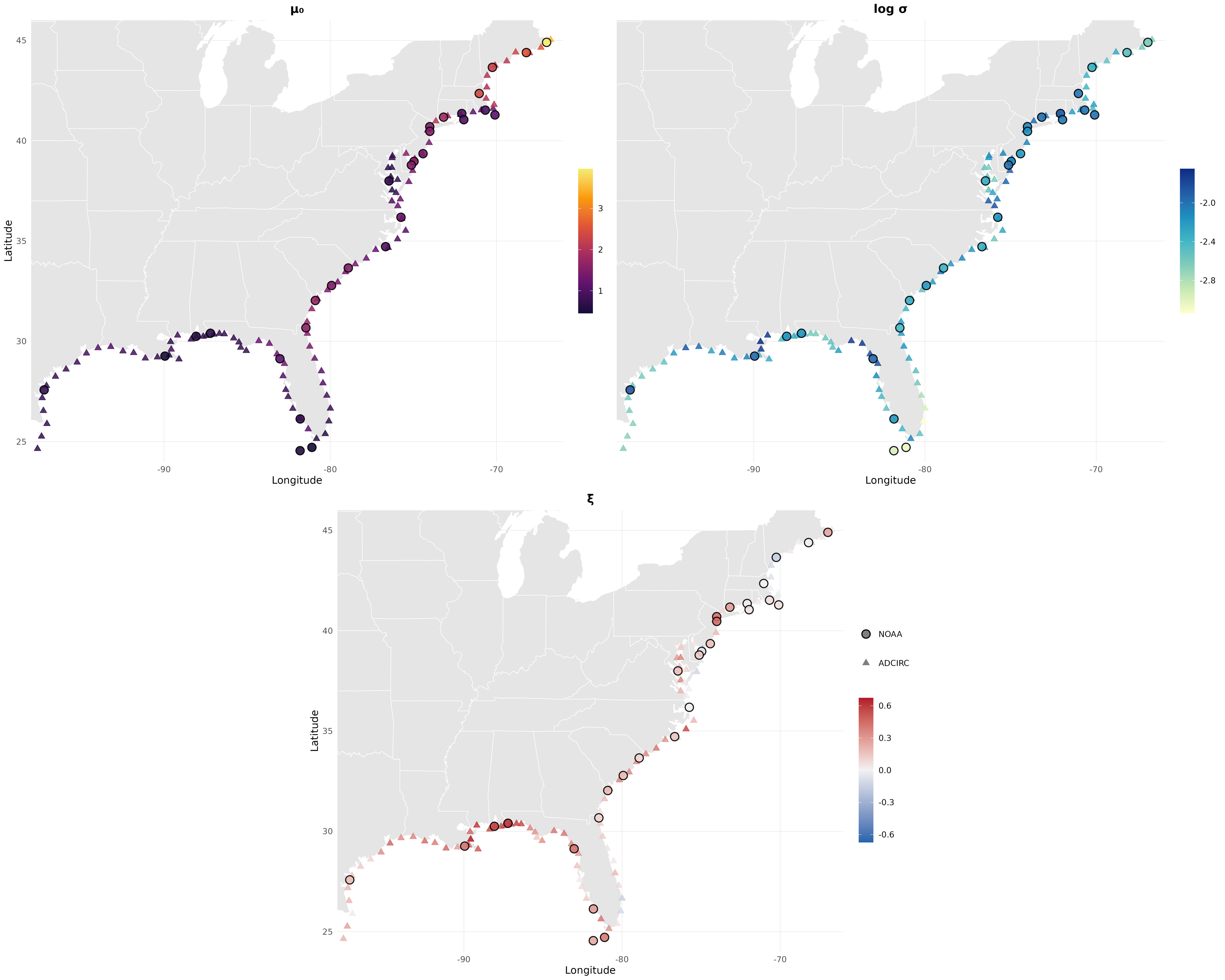}
\caption{Stage~1 pointwise GEV maximum likelihood estimates at 129 sites. NOAA tidal gauges (circles with outlines) and ADCIRC simulation sites (triangles). Left: location parameter (meters). At NOAA sites this is $\hat{\mu}_0$, the year-2000 intercept from the nonstationary fit; at ADCIRC sites this is the stationary MLE $\hat{\mu}$. Center: $\log\sigma$. Right: shape $\xi$, with diverging color scale centered at zero. The location parameter increases from south to north along the Atlantic coast, while the Gulf Coast exhibits elevated positive $\xi$ values indicating heavy tails.}
\label{fig:stage1}
\end{figure}

The estimated location parameter ($\hat{\mu}_0$ at NOAA, $\hat{\mu}$ at ADCIRC) ranges from 0.46 to 3.96\,m, with a strong south-to-north gradient along the Atlantic coast. The shape parameter $\hat{\xi}$ shows clear geographic structure: positive values ($\sim$0.2--0.5) along the Gulf Coast indicate heavy tails, while near-zero or slightly negative values north of Cape Hatteras indicate lighter tails. NOAA and ADCIRC estimates are broadly consistent, with ADCIRC sites filling gaps in the gauge network (Figure~\ref{fig:stage1}).

\subsection{Fitted Spatial Model}
\label{sec:fitted_model}

Multi-start optimization (20 starts) converges to a consistent optimum. The estimated cross-source correlations are:
\begin{equation}
\mathrm{Cor}(\mu_N, \mu_A) = 0.995, \qquad \mathrm{Cor}(\xi_N, \xi_A) = 0.837, \qquad \mathrm{Cor}(\log\sigma_N, \log\sigma_A) = 0.443.
\label{eq:correlations}
\end{equation}

The near-perfect location correlation ($r = 0.995$) is the primary driver of fusion: both sources capture the same spatial pattern, so the 100 ADCIRC sites effectively densify the 29-station gauge network. The shape correlation ($r = 0.84$) matters because $\xi$ is the least precisely estimated parameter yet has the largest influence on return levels. The moderate scale correlation ($r = 0.44$) suggests the two sources are less aligned in their scale parameters than in location or shape.

Estimated range parameters span an order of magnitude: $\hat{\rho}_1 = 2873$\,km (the dominant GP loading onto $\mu$) to $\hat{\rho}_4 = 188$\,km (fine-scale variability), supporting the non-separable LMC structure. The longest-range GP captures the large-scale latitude gradient in the location parameter; we prefer this over explicit geographic covariates in the mean function, which would require specifying a shared covariate response across sources. The lower-triangular constraint on $\bA$ resolves rotational non-identifiability: individual $\rho_i$ values depend on the ordering of components in $\btheta(s)$, but all predictions and cross-source correlations are invariant to this choice.

\begin{figure}[!htbp]
\centering
\includegraphics[width=7.5in,center]{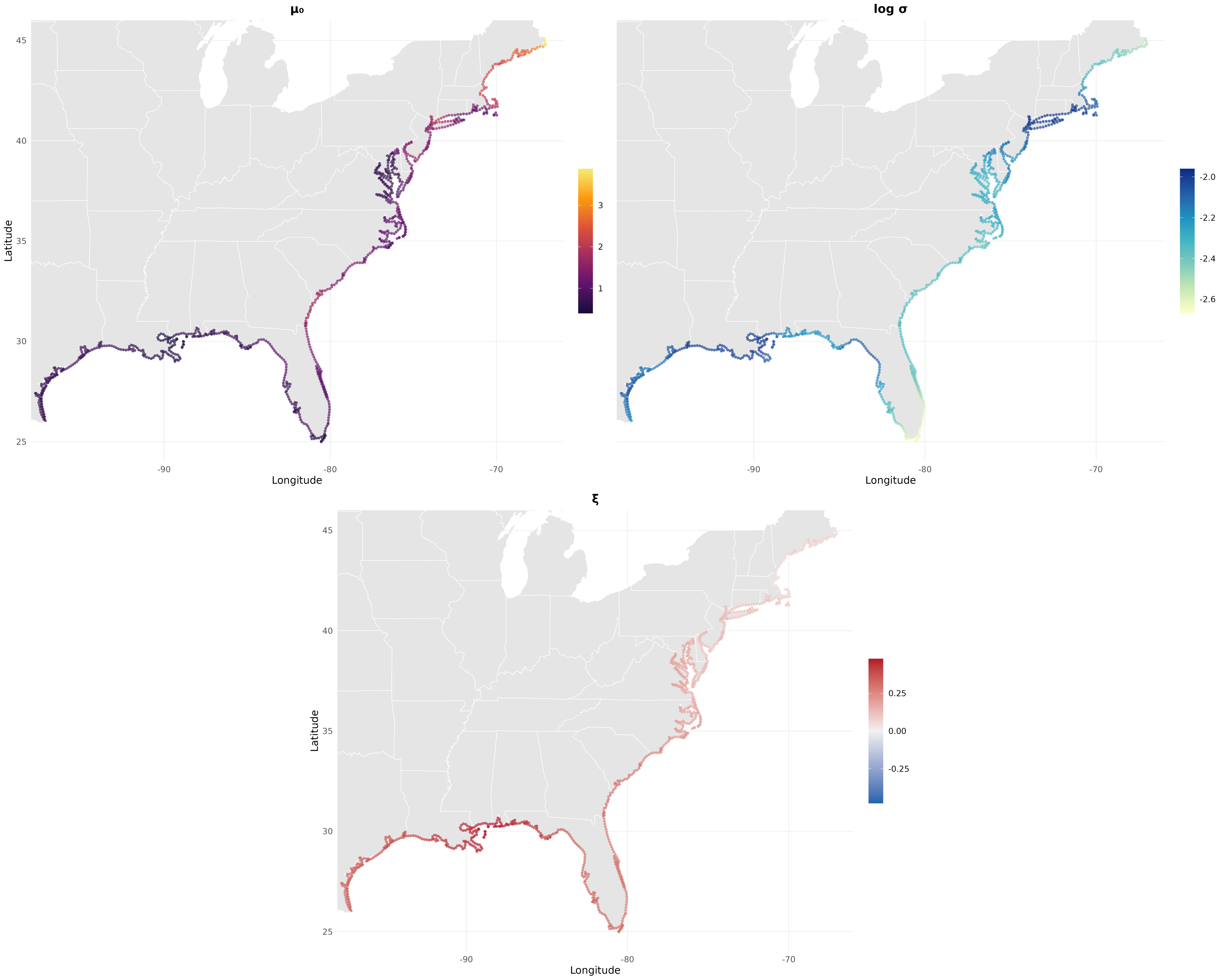}
\caption{Kriged NOAA-scale GEV parameters at the 29 NOAA sites from the joint model. Left: location $\mu_0$ (year-2000 intercept). Center: $\log\sigma$. Right: shape $\xi$, with diverging color scale centered at zero. Compare with the Stage~1 MLEs in Figure~\ref{fig:stage1}; the kriged fields are spatially smoother due to borrowing from ADCIRC.}
\label{fig:kriged}
\end{figure}

Figure~\ref{fig:kriged} shows the kriged NOAA-scale GEV parameters at the 29 gauge sites. Compared to the Stage~1 MLEs (Figure~\ref{fig:stage1}), the kriged fields are spatially smoother, reflecting the regularization from the joint spatial model. The location and shape parameters closely track the MLE pattern; the log-scale parameter shows the smallest departure from the Stage~1 values, consistent with the moderate cross-source correlation.

\subsection{Return Level Maps and Fusion Benefit}
\label{sec:return_levels}

\begin{figure}[!htbp]
\centering
\includegraphics[width=\textwidth]{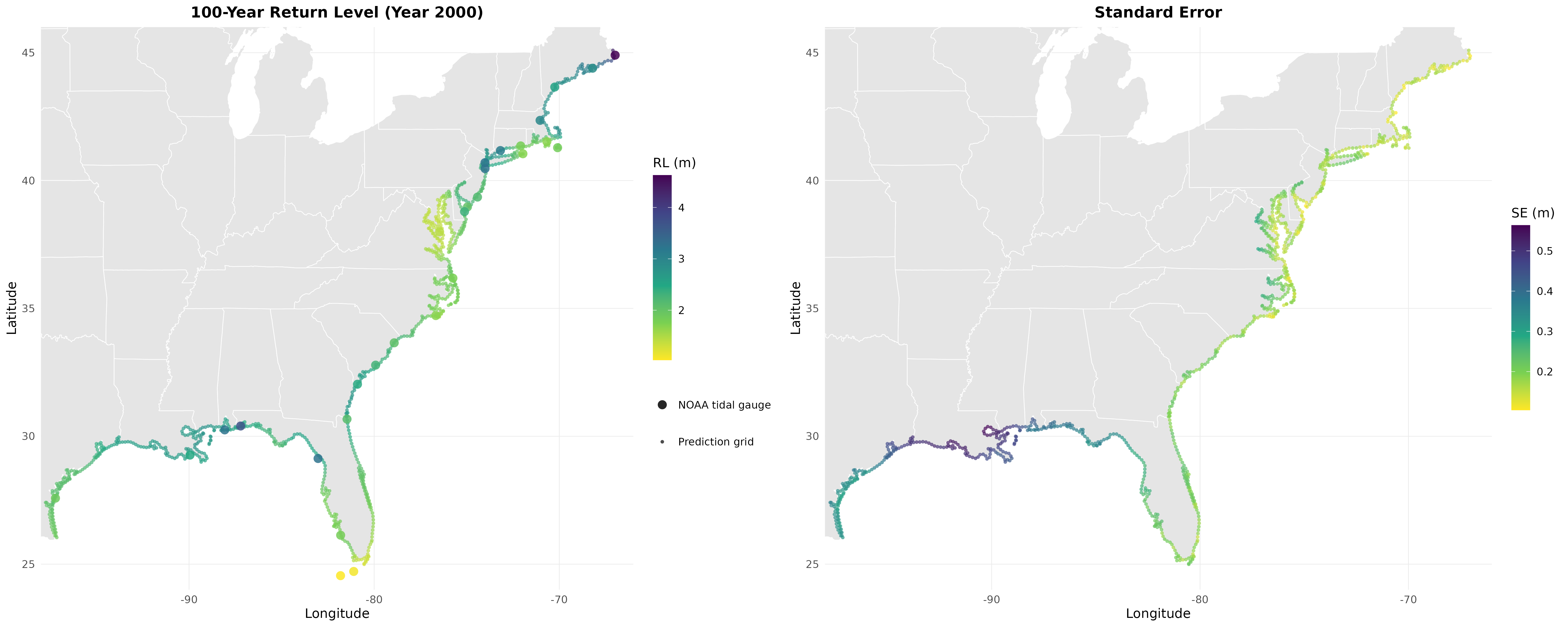}
\caption{Left: Kriged 100-year return levels (meters above MSL) at year-2000 reference conditions along the U.S.\ East and Gulf Coasts. Large circles show NOAA site-level MLE estimates; small points show joint-model predictions at a 15\,km coastal grid. Right: Standard errors of the 100-year return level (meters), via Monte Carlo simulation from the kriging posterior.}
\label{fig:rl100}
\end{figure}

Kriged 100-year return levels along a 15\,km coastal grid range from 1.28\,m in the Gulf of Mexico to 4.29\,m in northern New England (Figure~\ref{fig:rl100}, left), representing year-2000 reference conditions. Standard errors are lowest along the Atlantic coast (0.10--0.19\,m) where observations are densest, and largest along the Gulf Coast (0.16--0.54\,m) where sparser gauges and heavier tails compound uncertainty (Figure~\ref{fig:rl100}, right).

\begin{figure}[!htbp]
\centering
\includegraphics[width=\textwidth]{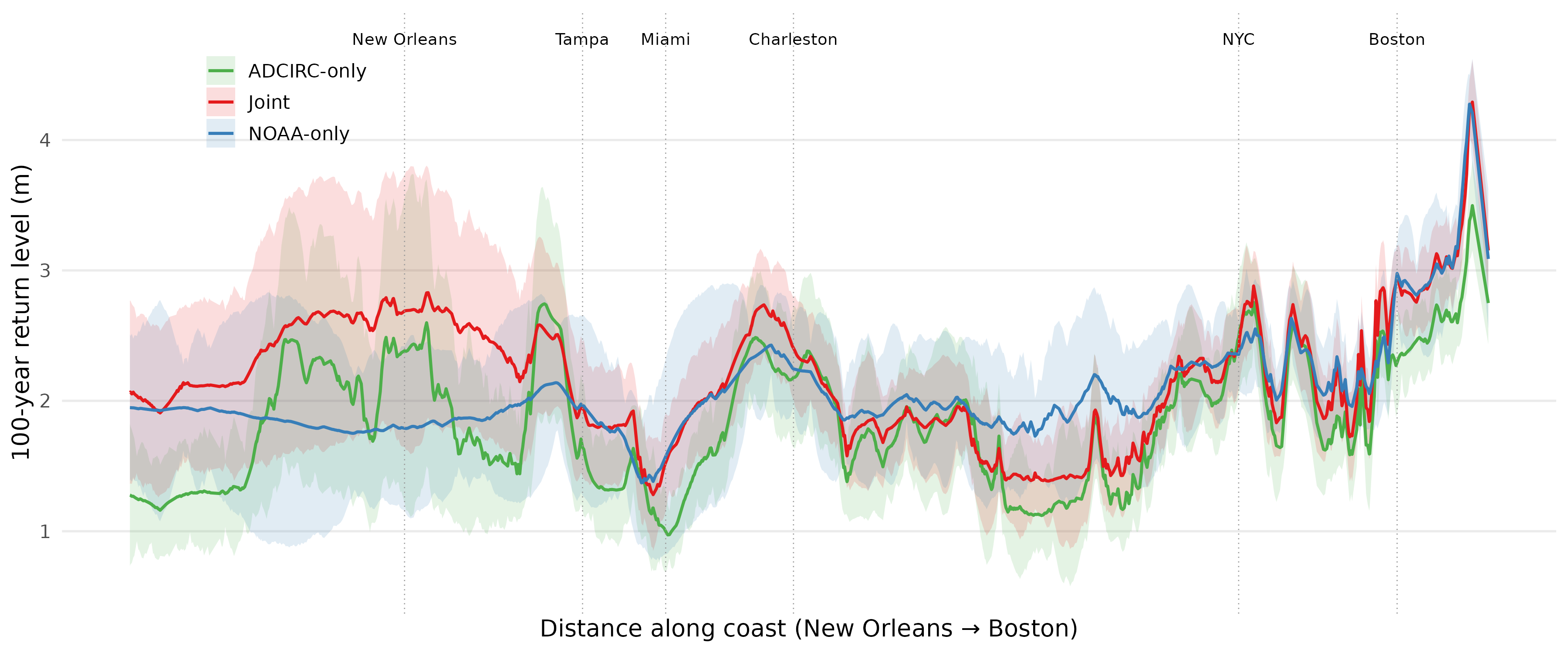}
\caption{100-year return level estimates at year-2000 reference conditions with 95\% confidence bands along the coast from New Orleans to Boston. The joint model (red) has narrower uncertainty than NOAA-only (blue) along the Atlantic coast; along the Gulf Coast the bands are comparable. The ADCIRC-only model (green) is systematically lower.}
\label{fig:rl_profile}
\end{figure}

\begin{figure}[!htbp]
\centering
\includegraphics[width=\textwidth]{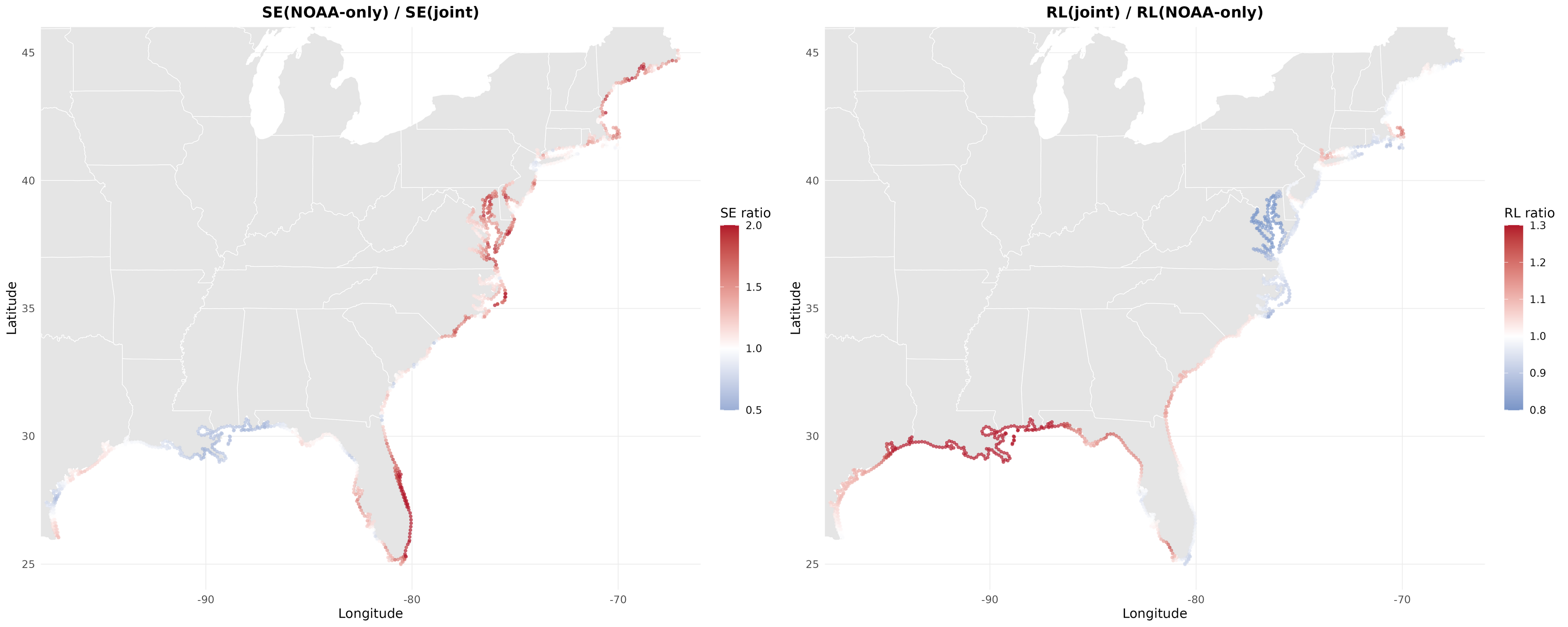}
\caption{Left: Ratio of 100-year return level standard errors (at year-2000 reference conditions; NOAA-only to joint model). Values greater than 1 indicate uncertainty reduction from fusion. Right: Ratio of 100-year return levels, joint model to NOAA-only. The largest uncertainty reductions are along the Southeast Atlantic; the largest point estimate shifts are along the Gulf Coast.}
\label{fig:ratio_maps}
\end{figure}

Comparing joint and NOAA-only models along the coast (Figure~\ref{fig:rl_profile}), fusion provides two distinct benefits whose relative importance varies geographically. Along the Atlantic coast, where ADCIRC coverage is densest, confidence bands narrow substantially (median SE ratio 1.34; Figure~\ref{fig:ratio_maps}, left). Along the Gulf Coast, SE ratios are close to 1 (median 0.92) and fusion instead shifts point estimates upward, as ADCIRC's heavy-tailed shape estimates pull $\xi$ higher where gauge coverage is sparse (Figure~\ref{fig:ratio_maps}, right). The median ratio of 100-year return level standard errors (NOAA-only to joint), hereafter the SE ratio, is 1.17, meaning the joint model's return level uncertainty is about 85\% of the NOAA-only value at a typical location. An SE ratio exceeding 1 indicates that fusion reduces uncertainty; roughly 24\% of the coastline shows SE ratios exceeding 1.5. The ADCIRC-only model is systematically lower than gauge-based estimates.

\subsection{Leave-One-Out Cross-Validation}
\label{sec:loocv}

Under LOO-CV at the 29 NOAA sites, the joint model reduces 100-year return level RMSE by 35\% (0.467\,m vs.\ 0.724\,m; Table~\ref{tab:loocv}), measured against each held-out site's Stage~1 MLE. The improvement is concentrated in the location parameter, consistent with the near-perfect cross-source correlation ($r = 0.995$). The shape parameter shows a modest 9\% reduction; the log-scale parameter shows no change (0.205 vs.\ 0.205), as expected given the moderate scale correlation. The total LPD gain is $+32.4$ across 29 sites; the joint model wins at 28 of 29 sites.

Parameter-level scatter plots (Figure~S5) confirm that the largest departures from the 1:1 line are in $\xi$, where ADCIRC pulls Gulf Coast estimates upward, and in $\log\sigma$, consistent with the moderate cross-source correlation ($r = 0.44$).

To isolate each parameter's contribution to the return level improvement, we replaced one parameter at a time in the NOAA-only LOO predictions with the joint model's LOO value and recomputed return levels (Table~\ref{tab:rmse_decomp}). The location parameter accounts for roughly half the RMSE reduction (18.1\%), the shape parameter contributes a further 10.5\%, and the log-scale parameter contributes nothing ($-1.0$\%). The combined effect (35.5\%) exceeds the sum of individual contributions because the parameters interact nonlinearly in the return level formula~\eqref{eq:rl}.

Probability integral transform (PIT) values assess calibration of the LOO predictive distributions (Table~\ref{tab:calibration}; PIT histograms in Figure~S4). Empirical coverage is at or above nominal levels: predictive intervals are wider than necessary, as expected because treating $\bW$ as fixed slightly overstates predictive variance. The $\mu$ and return level coverages are closest to nominal, while $\xi$ and $\log\sigma$ show overcoverage even at the 50\% level (24/29 and 23/29 vs.\ expected 14.5). Kolmogorov-Smirnov tests show no rejections ($p = 0.96$, $0.15$, $0.22$, $0.21$ for $\mu$, $\log\sigma$, $\xi$, and return levels), though power is limited with $n = 29$.

\subsection{Geographic Block Cross-Validation}
\label{sec:blockcv}

Table~\ref{tab:blockcv} summarizes the geographic block CV, in which each of five contiguous coastal blocks is held out and Stage~2 is refitted. Across all 29 sites, the joint model reduces 100-year return level RMSE by 18\% (0.665\,m vs.\ 0.810\,m). The most informative fold is the Gulf Coast: with all four gauges removed and the nearest retained NOAA site over 500\,km away, the NOAA-only model encounters near-singular kriging covariance and loses predictive capacity, while the joint model maintains reasonable predictions through ADCIRC coverage. Excluding the four Gulf Coast sites isolates the effect on the remaining 25 sites, where the joint model achieves a 36\% return level RMSE reduction (0.496\,m vs.\ 0.776\,m), confirming that fusion benefits persist under spatial extrapolation.

The log-scale parameter is the one case where fusion increases error: NOAA-only log-scale RMSE is 0.298 vs.\ 0.406 for the joint model. This is expected given the moderate cross-source scale correlation ($r = 0.44$): under spatial extrapolation, borrowing from ADCIRC scale parameters that differ substantially from NOAA values introduces noise rather than information. This parameter-specific result underscores that practitioners should examine per-parameter correlations before assuming uniform benefit from fusion.

\subsection{Sensitivity Analyses}
\label{sec:sensitivity}

\subsubsection{Simulation Network Density}
\label{sec:saturation}

\begin{figure}[!htbp]
\centering
\includegraphics[width=0.7\textwidth]{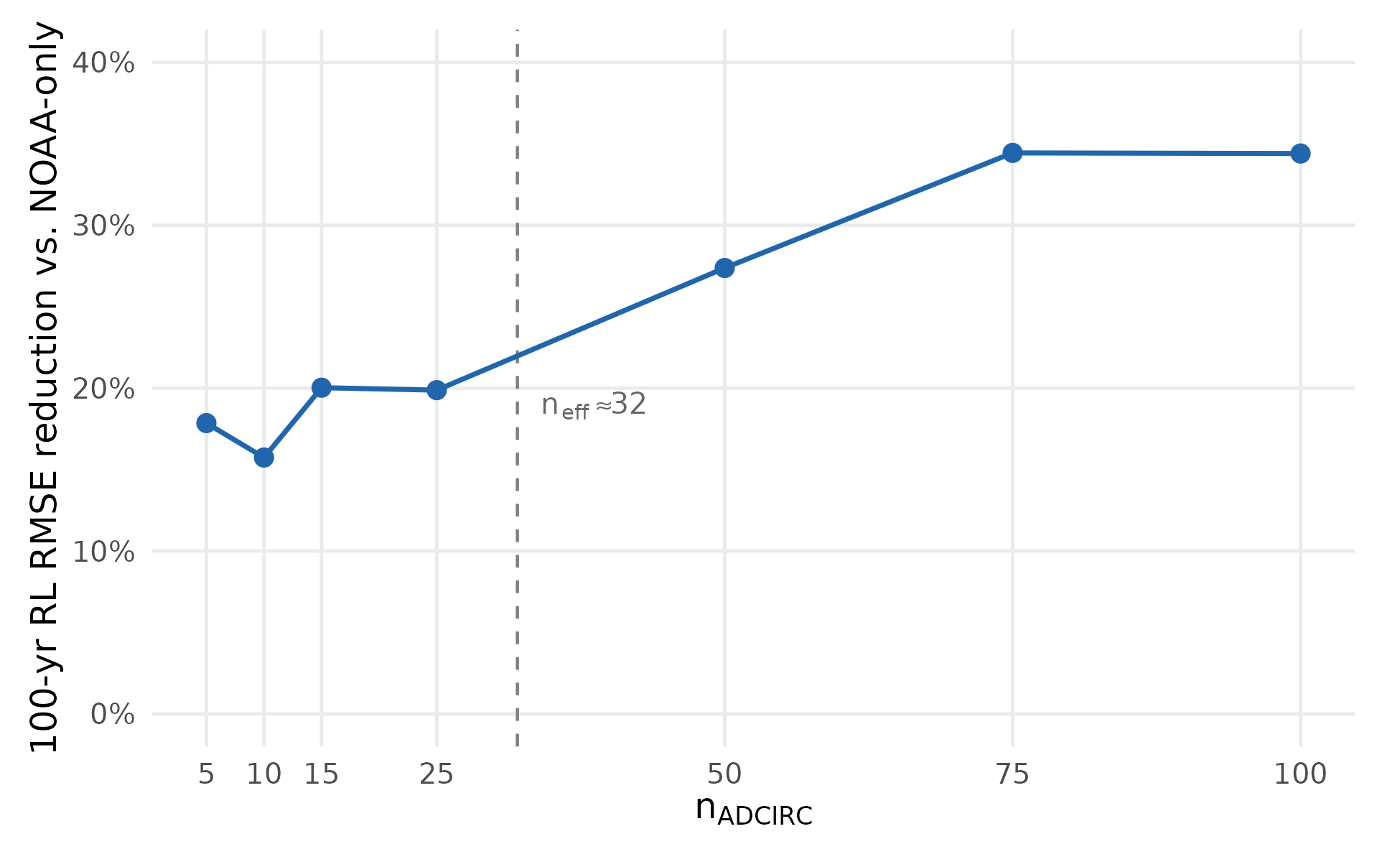}
\caption{Reduction in 100-year return level LOO-CV RMSE (relative to NOAA-only) as a function of the number of ADCIRC sites included. The dashed line marks $n_{\text{eff}} \approx 32$, approximately one site per minimum correlation range ($\hat{\rho}_{\min} = 188$\,km over $\sim$6000\,km of coastline), beyond which marginal gains diminish at the current spatial resolution. As few as 15 sites yield a 20\% reduction.}
\label{fig:saturation}
\end{figure}

The RMSE reduction saturates near 35\% once approximately 30 ADCIRC sites are included, roughly one per minimum correlation range ($\hat{\rho}_{\min} = 188$\,km over $\sim$6000\,km of coastline; Figure~\ref{fig:saturation}). As few as 15 sites yield a 20\% reduction. This threshold reflects the spatial resolution of the current analysis: large-scale coastline patterns are fully characterized by $\sim$30 sites over 6{,}000\,km. Analyses targeting finer-scale structure would require a denser draw from the ADCIRC grid and would likely raise the saturation threshold. This saturation analysis provides additional support for the 100-site ADCIRC network used in this study: the full benefit of fusion is captured well before 100 sites, suggesting that the current network is more than sufficient for the analysis.

\subsubsection{Taper Range and Analyst Choices}
\label{sec:taper}

Results are negligibly sensitive to the taper range over $\lambda \in \{150, 300, 500\}$\,km (Table~\ref{tab:taper_sensitivity}): the location correlation is stable ($r \approx 0.995$), return level RMSE is virtually identical (0.463--0.467\,m), and the LPD gain varies by about 10\%. The shape correlation varies modestly ($r = 0.81$--$0.90$). The exponential covariance was chosen for parsimony; 129 sites provide limited power to distinguish smoothness. Other choices (resampling unit, multi-start count, grid resolution) showed negligible sensitivity.

\subsubsection{Bootstrap Batch Stability}
\label{sec:batch_stability}

Splitting $B = 2{,}000$ bootstrap replicates into four independent batches, the location correlation is effectively invariant ($r = 0.994$--$0.996$), the shape correlation is stable ($r = 0.81$--$0.85$), and return level LOO-RMSE ranges from 0.454 to 0.491\,m. The scale correlation is least stable ($r = 0.44$--$0.78$), reflecting a flat likelihood surface; because the scale parameter contributes least to return level variation, this instability has minimal impact on predictions. Kriged return level maps are visually indistinguishable across batches.

\subsubsection{Identifiability under Non-co-location}
\label{sec:identifiability}

To assess whether the cross-source correlations are recoverable under the study's sampling design, we generated 100 synthetic datasets from the fitted model (true $\bbeta$, $\bA$, $\bm{\rho}$) with measurement error drawn from the fixed bootstrap estimate $\hat{\bW}$ (treating Stage~1 uncertainty as known), applying the same partial observation structure (29 sites observing parameters 1--3, 100 sites observing parameters 4--6). Stage~2 was refitted to each dataset (Table~\ref{tab:simulation}).

The location correlation is precisely recovered. The log-scale correlation is noisy but approximately unbiased. The shape correlation exhibits high estimation uncertainty: the reported point estimate of 0.84 lies within the distribution of recovered values but the likelihood surface for this quantity is effectively flat with 29 within-source observations for $\xi$. Despite this uncertainty, the LOO-CV gains are robust: the RMSE decomposition (Table~\ref{tab:rmse_decomp}) shows that the $\xi$ contribution (10.5\% reduction) is secondary to the precisely estimated $\mu$ correlation (18.1\%).

\subsubsection{Uncertainty Components}
\label{sec:uncertainty_components}

The reported standard errors use Monte Carlo simulation (10{,}000 draws from the kriging posterior), which omit uncertainty in $\bW$ and in the Stage~2 parameters $(\bA, \bm{\rho})$. For comparison, we also computed delta method standard errors via~\eqref{eq:delta_se}. Under LOO-CV, the delta method substantially underestimates uncertainty: holding out an entire site forces long-range extrapolation, producing a wide kriging posterior over which the nonlinear return level function~\eqref{eq:rl} is poorly approximated by a first-order Taylor expansion. On the prediction grid, where interpolation keeps the kriging posterior narrow, the two methods agree closely (Figure~S6). The empirical overcoverage in Table~\ref{tab:calibration}, where predictive intervals are wider than necessary at all nominal levels, provides indirect evidence that ignoring uncertainty in $\bW$ and the Stage~2 parameters does not produce confidence intervals that are too narrow.

\section{Discussion}
\label{sec:discussion}

The near-perfect cross-source location correlation ($r = 0.995$) drives the fusion benefit, effectively densifying the 29-station NOAA gauge network with 100 simulation sites. The geographic pattern of the benefit (point estimate shifts along the Gulf Coast versus uncertainty reduction along the Southeast Atlantic) reflects the interplay of network density and tail behavior.

\subsection{Comparison with Existing Methods}
\label{sec:comparison}

The main modeling contribution is extending \citet{Russell2020}'s single-source LMC to a multi-source setting. The cross-source covariance block provides the mechanism for information transfer, even without co-located data. A selection matrix handles the partial observation structure within the standard LMC likelihood, requiring no additional distributional assumptions. The formulation accommodates additional sources or co-located observations without structural modification.

The LMC's principal advantage over simpler alternatives (such as nearest-neighbor bias correction using ADCIRC sites) is that it models spatial correlations jointly across all six parameters and propagates uncertainty coherently into return level confidence intervals. Ad hoc corrections can shift point estimates but cannot produce calibrated prediction intervals, which are essential for risk-based decision making.

The return level RMSE reduction is measured against a frequentist LMC baseline fitted to the 29-station gauge network alone. Fully Bayesian hierarchical models for spatial extremes \citep{HuserWadsworth2019, ZhangShabyWadsworth2022, ZhangRisserWehner2024} represent the main methodological alternative, modeling dependence at the observation level rather than through a latent Gaussian process. These approaches are computationally intensive, requiring MCMC over hundreds of coupled parameters with a non-conjugate GEV likelihood, and remain parametric models whose assumptions are difficult to verify at this scale. Our two-stage structure, by contrast, makes Stage~1 diagnostics directly verifiable, avoids iterating between stages, and produces directly interpretable cross-source correlations. Adapting fully Bayesian spatial extremes models to the multi-source fusion setting with partial observations is an open challenge.

Other recent approaches to extreme sea level estimation at ungauged U.S.\ coastal sites, including max-stable processes fitted to tide gauge annual maxima \citep{Rashid2024} and spatiotemporal Bayesian hierarchical frameworks \citep{Morim2025}, rely entirely on observational data. The present framework is distinct in fusing observations with simulation output; the 35\% return level RMSE improvement over our gauge-only baseline quantifies this benefit.

\subsection{Simulator Bias and Model Discrepancy}
\label{sec:bias}

The LMC assumes that NOAA and ADCIRC GEV parameters are realizations of a common spatial process, differing by source-specific loadings on shared latent GPs but not by a systematic spatial bias field. ADCIRC-only return levels are systematically lower than gauge-based estimates (Figure~\ref{fig:rl_profile}). Within the LMC, this level difference is absorbed into the source-specific loadings in $\bA$, producing lower ADCIRC $\mu$ estimates without a separate bias term.

The critical assumption is that ADCIRC's spatial \emph{pattern} of GEV parameters is aligned with reality, even if its levels are shifted. The near-perfect location correlation ($r = 0.995$) supports this: both sources rank locations identically by extreme sea level. The LOO-CV provides an empirical check: if ADCIRC introduced spatially structured bias, the joint model would systematically mispredict at sites near ADCIRC clusters, but no such pattern is evident. A formal discrepancy model (e.g., source-specific additive GPs) would require co-located observations to separately identify the bias field and the cross-source covariance; where such data are available, the extension would be straightforward.

\subsection{Practical Recommendations}
\label{sec:recommendations}

The framework is most beneficial when (i) the ground-truth network is sparse relative to the spatial correlation range, (ii) the simulation source reproduces the spatial pattern of the quantity of interest even if its levels are biased, and (iii) cross-source correlations for the hardest-to-estimate parameters (typically $\xi$) are sufficiently high. In our application, the saturation analysis (Section~\ref{sec:sensitivity}) provides a practical rule of thumb: approximately one simulation site per minimum correlation range ($n_{\text{eff}} \approx L_{\text{domain}}/\rho_{\min}$, where $L_{\text{domain}}$ is the domain length) suffices to capture the full fusion benefit; beyond this, marginal gains diminish. As few as 15 ADCIRC sites yielded a 20\% return level RMSE reduction.

The spatial intermixing of the two networks matters more than their individual densities: inter-source distances should be well below $\rho_{\min}$ to identify the cross-source covariance block. Practitioners should examine per-parameter correlations before assuming that all components benefit equally; for the moderately correlated log-scale parameter ($r = 0.44$), borrowing from simulations increased rather than reduced prediction error under spatial extrapolation (Section~\ref{sec:blockcv}).

Beyond coastal sea levels, the same framework applies wherever sparse ground-truth observations coexist with dense model output: precipitation extremes with radar composites or reanalysis, or air quality monitoring with satellite retrievals. The selection matrix formalism extends naturally to settings with co-located observations, additional data sources, or asymmetric parameter dimensions across sources.

\subsection{Limitations and Future Work}
\label{sec:limitations}

Stage~2 assumes the Stage~1 MLEs are jointly Gaussian. For $\hat{\mu}_0$ and $\log\hat{\sigma}$ with 40+ annual maxima, this is well supported by asymptotic theory. For $\hat{\xi}$, convergence is slower near $\xi = 0$ \citep{Coles2001}, but Stage~1 diagnostics confirm adequate normality (Section~\ref{sec:stage1_results}).

The simulation study (Section~\ref{sec:identifiability}) reveals that the cross-source shape correlation, while positive, is estimated with high uncertainty under the present sampling design. The reported point estimate of $r = 0.84$ should be interpreted cautiously; the true value may be substantially lower. This does not undermine the return level gains, which are driven primarily by the precisely estimated location correlation, but it limits the strength of conclusions about cross-source tail dependence.

Several directions remain. Spatially varying trends at Stage~2 would enable return level projections at future reference years. The moderate scale correlation ($r = 0.44$) means ADCIRC contributes less information about $\log\sigma$ than about $\mu$ or $\xi$, likely reflecting genuine differences between the two sources in this parameter. The present analysis resolves large-scale coastline patterns with 100 ADCIRC sites over 6{,}000\,km. ADCIRC operates on unstructured grids with far finer resolution; exploiting this would enable sub-regional analyses but would require sparse likelihood approximations for the resulting high-dimensional covariance matrices, and potentially nonstationary covariance structure to capture spatially varying dependence. Even at the current resolution, fusion delivers a 35\% return level improvement, demonstrating that the framework is effective for broad-scale risk assessment with a computationally tractable analysis. A fully Bayesian formulation would eliminate the bootstrap covariance $\bW$ and propagate Stage~1 uncertainty automatically, but would require MCMC over approximately 420 coupled parameters with a non-conjugate GEV likelihood, and would produce a model that is much harder to interpret.

\section{Conclusion}
\label{sec:conclusion}

We have presented a two-stage frequentist framework for fusing sparse observations with dense simulations in spatial extreme value analysis, modeling GEV parameters from both sources as a single six-dimensional spatial process via the LMC. Applied to U.S.\ coastal sea levels with a nonstationary Stage~1 model, the method reduces 100-year return level LOO-CV RMSE by 35\% at year-2000 reference conditions, relative to a gauge-only model, with the benefit persisting under spatial extrapolation (18\% block CV reduction). The gain is driven by near-perfect cross-source location correlation and substantial shape correlation, manifesting as narrower confidence bands where gauge density is highest and as corrected point estimates where gauges are sparsest, though the moderately correlated scale parameter shows that fusion is not universally beneficial across all GEV components. The framework applies to any domain where sparse ground-truth observations coexist with dense model output.

\section*{Supporting Information}

Additional information and supporting material for this article is available online at the journal's website.

\section*{Data Accessibility Statement}

The R package \texttt{evfuse} implementing all methods is available at \url{https://github.com/briannathanwhite/evfuse}. The package bundles the processed annual maxima for all 129 sites (29 NOAA, 100 ADCIRC), which are sufficient to reproduce every result in this paper. Raw hourly sea-level observations are publicly available from the NOAA Tides \& Currents portal (\url{https://tidesandcurrents.noaa.gov}); raw ADCIRC simulation output was generated by the Renaissance Computing Institute (RENCI) at the University of North Carolina at Chapel Hill and is available from the authors upon request. Scripts to reproduce all analyses are included in the package repository.

\section*{Acknowledgments}

ADCIRC simulation data were provided by the Renaissance Computing Institute (RENCI) at the University of North Carolina at Chapel Hill. A preliminary version of this work was presented at the 14th International Conference on Extreme Value Analysis (EVA 2025), Chapel Hill, NC.

\section*{Funding}

No external funding was received for this work.

\section*{Competing Interests}

The authors declare no competing interests.

\bibliographystyle{apalike}

\clearpage

\begin{table}[p]
\centering
\caption{Trend diagnostic comparison between data sources.}
\label{tab:trends}
\smallskip
\begin{tabular}{@{}lcc@{}}
\toprule
& NOAA (observed) & ADCIRC (simulated) \\
\midrule
Sites tested & 29 & 100 \\
Significant (MK $p < 0.05$) & 19 (66\%) & 1 (1\%) \\
Median Sen's slope (mm/yr) & 4.6 & 0.4 \\
Slope range (mm/yr) & 2.0 to 13.7 & $-2.2$ to 5.3 \\
\bottomrule
\end{tabular}
\end{table}

\begin{table}[p]
\centering
\caption{Leave-one-out cross-validation at 29 NOAA sites (nonstationary Stage~1 model).}
\label{tab:loocv}
\smallskip
\begin{tabular}{@{}lccc@{}}
\toprule
Metric & Joint & NOAA-only & Reduction \\
\midrule
$\mu_0$ LOO-RMSE & 0.142 & 0.434 & 67.3\% \\
$\log\sigma$ LOO-RMSE & 0.205 & 0.205 & 0.0\% \\
$\xi$ LOO-RMSE & 0.147 & 0.161 & 8.7\% \\
100-yr RL LOO-RMSE (m) & 0.467 & 0.724 & 35.5\% \\
\midrule
Total LPD & $+23.84$ & $-8.53$ & \\
Sites joint wins (LPD) & \multicolumn{3}{c}{28 / 29 (97\%)} \\
\bottomrule
\end{tabular}
\end{table}

\begin{table}[p]
\centering
\caption{Decomposition of the 100-year return level LOO-CV RMSE reduction. Each ``hybrid'' row replaces only the named parameter with the joint model's LOO prediction, keeping other parameters at their NOAA-only values.}
\label{tab:rmse_decomp}
\smallskip
\begin{tabular}{@{}lcc@{}}
\toprule
Method & RL RMSE (m) & Reduction (\%) \\
\midrule
NOAA-only (all params) & 0.724 & -- \\
Hybrid: $\mu_0$ from joint & 0.594 & 18.1 \\
Hybrid: $\xi$ from joint & 0.648 & 10.5 \\
Hybrid: $\log\sigma$ from joint & 0.732 & $-1.0$ \\
Joint (all params) & 0.467 & 35.5 \\
\bottomrule
\end{tabular}
\end{table}

\begin{table}[p]
\centering
\caption{Empirical coverage of LOO predictive intervals at 29 NOAA sites (joint model). Values near the nominal level indicate well-calibrated uncertainty. The KS $p$-value tests uniformity of the probability integral transform.}
\label{tab:calibration}
\smallskip
\begin{tabular}{@{}lcccc@{}}
\toprule
Parameter & 50\% & 80\% & 95\% & KS $p$ \\
\midrule
$\mu$ & 17/29 & 23/29 & 28/29 & 0.962 \\
$\log\sigma$ & 23/29 & 27/29 & 28/29 & 0.145 \\
$\xi$ & 24/29 & 28/29 & 29/29 & 0.220 \\
100-yr RL & 19/29 & 26/29 & 29/29 & 0.212 \\
\midrule
\textit{Expected} & \textit{14.5} & \textit{23.2} & \textit{27.6} & \\
\bottomrule
\end{tabular}
\end{table}

\begin{table}[p]
\centering
\caption{Geographic block cross-validation at 29 NOAA sites. Five contiguous coastal blocks are held out in turn, with Stage~2 refitted for each fold. The ``excl.\ Gulf'' row excludes the four Gulf Coast sites where the NOAA-only model encounters near-singular kriging covariance, isolating the effect on the remaining 25 sites.}
\label{tab:blockcv}
\smallskip
\begin{tabular}{@{}lccc@{}}
\toprule
Metric & Joint & NOAA-only & Reduction \\
\midrule
$\mu_0$ RMSE & 0.255 & 0.781 & 67.3\% \\
$\log\sigma$ RMSE & 0.406 & 0.298 & $-36.2$\% \\
$\xi$ RMSE & 0.194 & 0.213 & 8.9\% \\
\midrule
100-yr RL RMSE, all 29 (m) & 0.665 & 0.810 & 17.9\% \\
100-yr RL RMSE, excl.\ Gulf (m) & 0.496 & 0.776 & 36.1\% \\
\midrule
Sites joint wins (LPD) & \multicolumn{3}{c}{24 / 29 (83\%)} \\
\bottomrule
\end{tabular}
\end{table}

\begin{table}[p]
\centering
\caption{Sensitivity of key results to the Wendland taper range $\lambda$ (km).}
\label{tab:taper_sensitivity}
\smallskip
\begin{tabular}{@{}lccc@{}}
\toprule
& $\lambda = 150$ & $\lambda = 300$ & $\lambda = 500$ \\
\midrule
$\mathrm{Cor}(\mu_N, \mu_A)$ & 0.995 & 0.995 & 0.994 \\
$\mathrm{Cor}(\log\sigma_N, \log\sigma_A)$ & 0.381 & 0.443 & 0.405 \\
$\mathrm{Cor}(\xi_N, \xi_A)$ & 0.899 & 0.837 & 0.812 \\
100-yr RL RMSE, Joint (m) & 0.463 & 0.467 & 0.467 \\
RL RMSE reduction (\%) & 36.1 & 35.5 & 35.5 \\
LPD gain (Joint $-$ NOAA) & 30.7 & 32.4 & 33.9 \\
\bottomrule
\end{tabular}
\end{table}

\begin{table}[p]
\centering
\caption{Recovery of cross-source correlations in 100 simulated datasets generated from the fitted model under the observed sampling design (no co-location).}
\label{tab:simulation}
\smallskip
\begin{tabular}{@{}lcccc@{}}
\toprule
Parameter & True & Median & SD & IQR \\
\midrule
$\mathrm{Cor}(\mu_N, \mu_A)$ & 0.995 & 0.989 & 0.014 & 0.014 \\
$\mathrm{Cor}(\log\sigma_N, \log\sigma_A)$ & 0.443 & 0.466 & 0.329 & 0.497 \\
$\mathrm{Cor}(\xi_N, \xi_A)$ & 0.837 & 0.448 & 0.451 & 0.695 \\
\bottomrule
\end{tabular}
\end{table}

\end{document}